\documentclass[a4paper,fleqn]{cas-sc}

\usepackage[numbers,sort&compress]{natbib}

\usepackage{lineno}
\usepackage{bm}
\usepackage{xcolor}
\usepackage{amssymb}
\usepackage{titlesec}
\usepackage[euler]{textgreek}
\usepackage[nodots]{numcompress}

\def\tsc#1{\csdef{#1}{\textsc{\lowercase{#1}}\xspace}}
\tsc{WGM}
\tsc{QE}
\tsc{EP}
\tsc{PMS}
\tsc{BEC}
\tsc{DE}
\begin{document}
\renewcommand{\figurename}{Fig.}

\titleformat{\section}[block]{\large\bfseries}{\arabic{section}}{1em}{}
\let\printorcid\relax
\let\WriteBookmarks\relax
\def\floatpagepagefraction{1}
\def\textpagefraction{.001}

\shorttitle{}

\shortauthors{Zhao et~al.}

\title [mode = title]{Modelling lined rock caverns subject to hydrogen embrittlement and cyclic pressurisation in fractured rock masses}
%\tnotemark[1,2]

%\tnotetext[1]{This document is t results of the research
%   project funded by the National Science Foundation.}

%\tnotetext[2]{The second title footnote which is a longer text matter
%   to fill through the whole text width and overflow into
%   another line in the footnotes area of the first page.}

\author[1,2]{Chenxi Zhao}

%\cormark[1]



%\credit{Conceptualization of this study, Methodology, Software}

\affiliation[inst1]{organization={Department of Geotechnical Engineering, College of Civil Engineering},%Department and Organization
            addressline={Tongji University}, 
            city={Shanghai},
            ostcode={200092}, 
            %state={},
            country={China}}

\affiliation[inst2]{organization={Key Laboratory of Geotechnical and Underground Engineering, Ministry of Education},%Department and Organization
            addressline={Tongji University}, 
            city={Shanghai},
            postcode={200092}, 
            %state={},
            country={China}}

 \affiliation[inst3]{organization={Department of Materials Science and Engineering, Uppsala University},%Department and Organization
            city={Uppsala},
            postcode={75236}, 
            country={Sweden}}

 \affiliation[inst4]{organization={Department of Earth Sciences, Uppsala University},%Department and Organization
            city={Uppsala},
            postcode={75236}, 
            country={Sweden}}
            
\author[3]{Haiyang Yu}
\author[1,2]{Zixin Zhang}

\author[4]{Qinghua Lei} [orcid=0000-0002-3990-4707]
\fnmark[*]
\ead{qinghua.lei@geo.uu.se}

\cortext[cor1]{Corresponding author}

%\fntext[fn1]{This is the first author footnote. but is common to third
%  author as well.}
%\fntext[fn2]{Another author footnote, this is a very long footnote and
%  it should be a really long footnote. But this footnote is not yet
%  sufficiently long enough to make two lines of %footnote text.}

%\nonumnote{This note has no numbers. In this work we demonstrate $a_b$
%  the formation Y\_1 of a new type of polariton on the interface
%  between a cuprous oxide slab and a polystyrene micro-sphere placed
%  on the slab.
%  }

%\linenumbers
\begin{abstract}
The technology of lined rock cavern (LRC) with great geographical flexibility is a promising, cost-effective solution to underground hydrogen storage. However, the air-tight steel tanks used in this technology are susceptible to material degradation due to hydrogen embrittlement (HE), potentially leading to leakage and structural failure, especial for LRCs constructed in complex geological conditions. In this paper, we develop a 2D multiscale numerical model based on the finite element method to assess the impact of HE on the LRC performance in fractured rock masses under cyclic gas pressurisation. Within this framework, a large-scale model is used to simulate the deformation and damage evolution of both fractured rock and an LRC under in-situ stresses and internal gas pressurisation, while a small-scale model captures HE in the steel lining of the LRC. Our simulations reveal that damage in the rock, concrete, and steel degradation is strongly affected by pre-existing fractures and in-situ stresses. Our results also reveal the presence of a strong positive feedback between hydrogen concentration and stress redistribution in the steel lining. Moreover, a comparison between models with and without considering HE illuminates that hydrogen concentration significantly contributes to steel degradation, particularly during the long-term LRC operation, highlighting the critical role of HE in the safety and performance of the LRC. The findings and insights obtained from our work have important implications for the design optimisation and performance assessment of LRCs for sustainable underground hydrogen storage.
\end{abstract}

% \includegraphics{figs/grabs.pdf}

%\item A multiscale model is developed to simulate lined rock caverns for hydrogen storage.
%\item Interaction between the cavern and its surrounding fractured rock mass is captured.
%\item Effects of hydrogen diffusion and embrittlement in steel linings are considered.
%\item Pre-existing fractures in rock exert a strong control on the damage in concrete and steel linings.
%\item Steel degradation is driven by the interplay of hydrogen concentration and stress distribution.

% Keywords

\maketitle

\section{Introduction}
\label{sec:Introduction}
The escalating impacts of climate change call for urgent and transformative strategies to counter global warming. Achieving the ambitious goal of net-zero carbon emissions by 2050 requires a fundamental shift in industrial practices, particularly in transitioning away from the use of fossil fuels. Among different potential solutions, hydrogen has emerged as a viable energy carrier, offering a sustainable alternative to traditional fossil fuels, by storing energy generated from weather-dependent intermittent renewables like wind, tidal, and solar. More specifically, techniques like electrolysis can be employed to convert excess renewable electricity during off-peak periods into hydrogen, which can then be stored underground for efficient and sustainable energy use. This strategy is widely regarded as a cornerstone of future renewable energy systems \citep{Tarkowski2019,TARKOWSKI202120010,Krevor2023,Schultz2023a,Schultz2023b}.

Several options for underground hydrogen storage have been proposed, including depleted oil and gas reservoirs, saline aquifers, and salt caverns as well as unlined or lined rock caverns \citep{Tarkowski2019}. However, each option is faced with specific challenges. Depleted hydrocarbon reservoirs and saline aquifers, for instance, are associated with risks of induced seismicity and potential hydrogen loss due to leakage or microbial activity \citep{Heinemann2021,Krevor2023}. Salt caverns, while effective, are geographically limited due to the scarcity of suitable salt deposits. Unlined rock caverns require substantial burial depths to achieve hydrodynamic sealing, making them costly to construct and maintain \citep{Lindblom1985}. In contrast, lined rock caverns (LRCs) offer a very flexible and cost-effective solution, as they can be constructed at shallower depths and are less geographically constrained \citep{Johansson2018,Patanwar2024}.

The development of LRCs for underground gas storage has a long history. Initial research began in 1985, leading to the construction of a pilot plant in Grängesberg, Sweden, during the late 1980s to experimentally test this concept \citep{Johansson2003}. This was followed by the establishment of a demonstration plant in Skallen, Sweden, between 1998 and 2002, which has remained operational until today \citep{Johansson2003,Glamheden2006}. These projects have generated extensive experiences and datasets, providing critical insights into the feasibility and performance of LRCs. More recently, the Hydrogen Breakthrough Ironmaking Technology (HYBRIT) initiative, launched in 2016, has integrated LRC-based hydrogen storage as a key component of its fossil-free steelmaking process. As part of this initiative, a pilot LRC facility has been constructed and commissioned in Luleå, Sweden, since the summer of 2022 \citep{Backlin2022,Vattenfall2022}. This facility represents a significant step forward in the practical application of hydrogen storage technologies.

An LRC system is composed of multiple components, each serving for a distinct purpose \citep{Johansson2003,Damasceno2023a,Damasceno2023b,LEONG2025749}. From the innermost to the outermost, these components include a steel lining, a bitumen sliding layer, a reinforced concrete lining, a shotcrete layer equipped with drainage pipes, and the surrounding rock mass. The stability and functionality of an LRC depend heavily on the load-bearing capacity of the rock mass and the reinforced concrete lining, while the steel lining primarily ensures gas containment. When the LRC is filled with gas, the resulting internal gas pressure causes volumetric expansion of the cavern, which can potentially lead to cracking in concrete or yielding in steel, compromising the safety and operational performance of the LRC.

Over the past decades, significant research efforts have been devoted to evaluating the performance of LRCs under high-pressure gas storage conditions through numerical simulations, experimental tests, and analytical calculations. For instance, \cite{Lu1998} constructed a 3D finite element model to simulate the deformation of an LRC during the pilot tests in Grängesberg, with results aligned closely with field measurements. Similarly, \cite{Glamheden2006} utilised a 2D finite difference model to study cavern convergence during the excavation phase of the LRC demonstration plant at Skallen. Additionally, coupled thermo-hydro-mechanical simulations have been employed to assess the mechanical stability, gas tightness, and energy efficiency of LRCs for compressed air energy storage \citep{Kim2012,Kim2013,Kim2016,Rutqvist2012}. Experimental studies have also played a crucial role in understanding the LRC behaviour. For example, \cite{Tunsakul2013} and \cite{Jongpradist2015} conducted laboratory experiments using physical models to study fracture growth in surrounding rocks induced by cavern pressurisation. Building on these findings, \cite{Tunsakul2014,Tunsakul2018} applied an element-free Galerkin method to simulate fracture propagation, validating their modelling results against experimental observations. Furthermore, \cite{Perazzelli2016} used a finite difference code to evaluate the resilience of LRCs against various failure scenarios, including rock mass uplift, steel lining buckling, fatigue, and concrete plug instability. Recent advancements in modeling techniques have further enhanced the understanding of LRC performance. \cite{Damasceno2023a} developed a 2D finite element model to analyse the interactions among the different LRC components, including the steel lining, sliding layer, reinforced concrete lining, and the surrounding jointed rock mass. In a separate study, \cite{Damasceno2023b} combined analytical models with 2D and 3D finite element simulations to estimate tangential strain on the cavern wall under high internal gas pressure. From a design perspective, some new approaches have also been proposed to optimise LRC construction. \cite{Park2013} introduced a probability-based design methodology that integrates point estimate methods with numerical modelling. More recently, \cite{Damasceno2023c} developed a reliability-based design tool using adaptive directional importance sampling coupled with 3D finite element analysis, offering a robust framework for LRC design and performance assessment. To better understand the impact of pre-existing rock fractures on the LRC behaviour, \cite{ZHAO2025252} developed a finite element model to study the deformation and damage evolution of LRCs and their interactions with surrounding fractured rock masses. Their results indicate that natural fractures tend to exert profound influences on the structural deformation of LRCs.

To apply the LRC technology for underground hydrogen storage, some new challenges need to be addressed. For example, under high gas pressure within the cavern, hydrogen could diffuse into the steel lining that is in direct contact with the hydrogen gas, leading to hydrogen embrittlement (HE) \citep{Molavitabrizi2022,Patanwar2024,Yu2024} and strength degradation of steel materials. This can result in the failure of steel linings and therefore loss of gas-tight integrity of the LRC system. With the development of steel degradation, changes in the local strain field can alter the pattern of hydrogen concentration, manifesting a coupling between steel degradation and hydrogen diffusion. Furthermore, for LRCs situated in fractured rock masses, it is expected that natural fractures may lead to stress concentrations in the LRC linings, potentially leading to localised hydrogen accumulation and embrittlement as well as increased risk of hydrogen leakage from the container. Apart from that, natural fractures often strongly control the overall geomechanical behavior of rock masseses \citep{Rutqvist2002,Barton2015,Lei2017a} and therefore could significantly impact the deformational response of LRC structures \citep{ZHAO2025252,QIU2024605}, which should be considered when assessing the performance of LRCs for underground hydrogen storage.

In this paper, we present a 2D multiscale numerical model to investigate the impact of HE on the performance of an LRC subject to cyclic hydrogen gas pressurisation and interaction with surrounding fractured rock masses. This multiscale model comprises a large-scale model for simulating deformation/damage evolution in both the LRC and rock masses and a small-scale model for capturing HE in the LRC's steel lining. This framework is capable of modelling various key processes across spatiotemporal scales in the system, such as fracture deformation and propagation in rock masses, damage evolution in concrete, and strength degradation in steel as well as their couplings. The remainder of the paper is organised as follows. Section \ref{sec:consititutive} introduces the governing equations for modelling deformation and damage in fractured rock and concrete as well as hydrogen diffusion and embrittlement in steel. Section \ref{sec:Model setup} describes the model setup including fracture network generation, model parameterisation, and mesh discretisation. Section \ref{sec:simulation results} presents and analyses simulation results, followed by a further discussion in Section \ref{sec:Discussion}. Section \ref{sec:conslusions} concludes the paper.

\section{Governing equations}
\label{sec:consititutive}

This section begins with an overview of the damage model designed to simulate crack propagation in rock and concrete materials in the large-scale model, which is complemented by a detailed explanation of the constitutive laws governing the displacement behaviour of pre-existing fractures in rock. Subsequently, we introduce the equations governing hydrogen diffusion in steel and the constitutive equations for capturing HE in the small-scale model. Our framework is developed based on the computational platform of COMSOL Multiphysics \citep{COMSOL2018}. Throughout the paper, we adopt the solid mechanics sign convention with tensile stresses/strains being positive and compressive stresses/strains negative.

\subsection{Solid deformation and damage evolution}
\label{sec:Damage}

The mechanical equilibrium of solids is governed by:
\begin{equation}
    \nabla \cdot {\bm{\sigma}} = 0,
\end{equation}
where $\nabla$ represents the gradient of a scalar field and $\bm{\sigma}$ is the stress tensor. A scalar-based continuum damage constitutive model is utilised to simulate fracturing in brittle/semi-brittle materials like rock and concrete here \citep{Jirasek2012}, with the constitutive equation given by:
\begin{equation}
\bm{\sigma} = (1-\omega)\boldsymbol{D} : \bm{\varepsilon},
\end{equation}
where $\omega$ ranging from 0 to 1 is the damage coefficient, $\boldsymbol{D}$ is the elastic stiffness matrix, and $\bm{\varepsilon}$ is the strain tensor. The progression of damage is governed by the following loading-unloading conditions \citep{Jirasek2012}:
\begin{equation}
f(\tilde{\varepsilon},\kappa)\leq 0,\dot{\kappa}\geq 0,\dot{\kappa}f(\tilde{\varepsilon},\kappa)=0, \label{xx}
\end{equation}
where $\tilde{\varepsilon}$ is the equivalent strain and $\kappa$ is an internal variable recording the historical maximum equivalent strain. We employ an elasto-brittle model to simulate rock damage \citep{Lei2021b,Zhao2021}, while concrete damage is simulated using the Mazars model \citep{Mazars2022}. The mathematical formulations of both models are described in the following.

\subsubsection{Damage in rock}
The smooth Rankine criterion is employed to compute the equivalent tensile and compressive strains in rock (denoted as $\tilde{\varepsilon}_\mathrm{rt}$ and $\tilde{\varepsilon}_\mathrm{rc}$, respectively) \citep{Jirasek2012}, given as:
\begin{equation}
\tilde{\varepsilon}_\mathrm{rt}=\frac{\| \langle \boldsymbol{D : \bm{\varepsilon}}\rangle_+\|}{E_\mathrm{r}}\textcolor{black}{\  ,} \label{xxa}
\end{equation}
and
\begin{equation}
\tilde{\varepsilon}_\mathrm{rc}=-\frac{\| \langle \boldsymbol{-D : \bm{\varepsilon}}\rangle_+\|}{E_\mathrm{r}}\textcolor{black}{\  ,}
\end{equation}
where $\| \cdot \|$ is the norm operator, $\langle \cdot \rangle_+$ are the Macaulay brackets denoting the positive part in the tensor, and $E_\mathrm{r}$ is the Young's modulus of the rock matrix.

When the tensile (or compressive) stress exceeds the material's tensile (respectively compressive) strength, a transition from elastic behavior to brittle failure would occur, associated with an abrupt stress drop (see Fig. \ref{fig:constitutive}a and b). The tensile and compressive damage variables (respectively denoted as $\omega_\mathrm{rt}$ and $\omega_\mathrm{rc}$, respectively) can be calculated as \citep{Lei2021b,Zhao2021}:
\begin{equation}
\omega_{\mathrm{rt}}=\left\{
\begin{split}
0 \qquad, &\ \kappa_\mathrm{t} \textless \varepsilon_\mathrm{rt0} \\
1-\frac{f_\mathrm{rtr}}{E\kappa_\mathrm{t}}, &\ \kappa_\mathrm{t} \geqslant \varepsilon_\mathrm{rt0}
\textcolor{black}{\  ,}
\end{split}
\right.
\end{equation}
and
\begin{equation}
\omega_{\mathrm{rc}}=\left\{
\begin{split}
0 \qquad, &\ \kappa_\mathrm{c} \textgreater \varepsilon_\mathrm{rc0} \\
1-\frac{f_\mathrm{rcr}}{E\kappa_\mathrm{c}}, &\ \kappa_\mathrm{c} \leqslant \varepsilon_\mathrm{rc0}\textcolor{black}{\  ,}
\end{split}
\right.
\end{equation}
where $\varepsilon_\mathrm{rt0}=f_\mathrm{rt0}/E_\mathrm{r}$ and $\varepsilon_\mathrm{rc0}=-f_\mathrm{rc0}/E_\mathrm{r}$ are the elastic limits of tensile and compressive strains, respectively; $f_\mathrm{rt0}$ and $f_\mathrm{rc0}$ represent the corresponding peak tensile and compressive strengths, while residual tensile and compressive strengths are calculated as $f_\mathrm{rtr} = \eta f_\mathrm{rt0}$ and $f_\mathrm{rcr} = \eta f_\mathrm{rc0}$, with $\eta$ being the residual strength ratio; $\kappa_\mathrm{t}$ and $\kappa_\mathrm{c}$ denote the maximum historical strains for tensile and compressive scenarios, respectively.

\subsubsection{Damage in concrete}
\label{sec:concrete damage}

Damage in concrete, denoted by $\omega_\mathrm{c}$, results from the co-evolution and interaction of tensile and compressive damage, which is approximated as \citep{Mazars2022}:
\begin{equation}
\omega_\mathrm{c} = \alpha_\mathrm{t} \omega_\mathrm{ct} + \alpha_\mathrm{c} \omega_\mathrm{cc}\textcolor{black}{\  ,}
\end{equation}
where the weighting coefficients $\alpha_\mathrm{t}$ and $\alpha_\mathrm{c}$ determine the relative intensity of tension and compression, respectively; $\omega_\mathrm{ct}$ and $\omega_\mathrm{cc}$ represent the tensile and compressive damage variables for concrete. These parameters can be calculated as follows. First, the stress and strain partitions are determined in the principal coordinate system as:
\begin{equation}
\left\{
\begin{split}
&\bm{\sigma} = \langle \bm{\sigma} \rangle_+ + \langle \bm{\sigma} \rangle_- \\
&\bm{\varepsilon} = \bm{\varepsilon}_\mathrm{t} + \bm{\varepsilon}_\mathrm{c}\textcolor{black}{\  ,}
\end{split}
\right.
\end{equation}
where 
\begin{equation}
\left\{
\begin{split}
&\bm{\varepsilon}_\mathrm{t} = \frac{1+\nu}{E} \langle \bm{\sigma} \rangle_+ - \frac{\nu}{E} \mathrm{tr} \left(\langle \bm{\sigma} \rangle_+\right)\\
&\bm{\varepsilon}_\mathrm{c} = \frac{1+\nu}{E} \langle \bm{\sigma} \rangle_- - \frac{\nu}{E} \mathrm{tr} \left(\langle \bm{\sigma} \rangle_-\right)\textcolor{black}{\  ,}
\end{split}
\right.
\end{equation}
where $\langle \cdot \rangle_+$ and $\langle \cdot \rangle_-$ represent the positive and negative parts of the tensor, respectively. The weighting coefficients are calculated as \citep{Mazars2022}:
\begin{equation}
\left\{
\begin{split}
&\alpha_\mathrm{t} = \Sigma_i H_i \frac{\varepsilon_{\mathrm{t}i} (\varepsilon_{\mathrm{t}i}+ \varepsilon_{\mathrm{c}i})}{\tilde\varepsilon_\mathrm{c}^2}\\
&\alpha_\mathrm{c} = \Sigma_i H_i \frac{\varepsilon_{\mathrm{c}i} (\varepsilon_{\mathrm{t}i} + \varepsilon_{\mathrm{c}i})}{\tilde\varepsilon_\mathrm{c}^2}\textcolor{black}{\  ,}
\end{split}
\right.
\end{equation}
where $i = 1, 2, 3$ represents the index of principal components; $H_{i}$ is the Heaviside step function, which takes the value of 1 when $\varepsilon_{i} > 0$, and 0 when $\varepsilon_{i} \leqslant 0$; $\tilde\varepsilon_\mathrm{c}$ is the Mazars equivalent strain, given by \citep{Mazars2022}:
\begin{equation}
\tilde\varepsilon_\mathrm{c} = \| \langle \mathbf{\bm{\varepsilon}}\rangle_+\|.
\end{equation}
Tensile and compressive damage variables can then be computed as follows \citep{Mazars2022}:
\begin{equation}
\left\{
\begin{split}
&\omega_\mathrm{ct} = 1-\frac{(1-A_\mathrm{t})\varepsilon_\mathrm{ct0}}{\mathrm{max}(\kappa_\mathrm{t},\varepsilon_\mathrm{ct0})} - A_\mathrm{t} \mathrm{exp}[-B_\mathrm{t}(\kappa-\varepsilon_\mathrm{ct0})]\\
&\omega_\mathrm{cc} = 1-\frac{(1-A_\mathrm{c})\varepsilon_\mathrm{cc0}}{\mathrm{max}(\kappa_\mathrm{c},\varepsilon_\mathrm{cc0})} - A_\mathrm{c} \mathrm{exp}[-B_\mathrm{c}(\kappa-\varepsilon_\mathrm{cc0})]\textcolor{black}{\  ,}
\end{split}
\right.
\end{equation}
where $A_\mathrm{t}, A_\mathrm{c}, B_\mathrm{t}$ and $B_\mathrm{c}$ are material parameters that influence the shape of post-peak response; $\varepsilon_\mathrm{ct0}=f_\mathrm{ct0}/E_\mathrm{c}$ and $\varepsilon_\mathrm{cc0}=-f_\mathrm{cc0}/E_\mathrm{c}$ represent the elastic limits of tensile and compressive strains, respectively, with $E_\mathrm{c}$ denoting the Young's modulus of the concrete.

\subsection{Fracture normal and shear deformations}
\label{sec:fracture constitutive}

\begin{figure}[!htb]
\centering
\includegraphics[width = 14 cm]{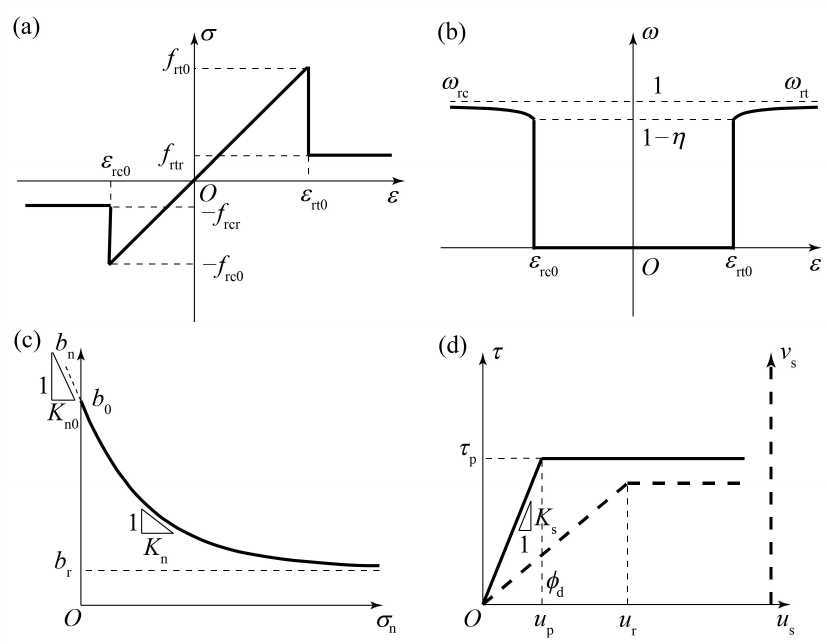}
\par
\caption{\textcolor{black}{Constitutive laws for rock matrix and pre-existing fractures\textcolor{black}{.} (a) \textcolor{black}{S}tress\textcolor{black}{–}strain\textcolor{black}{,} (b) damage\textcolor{black}{–}strain relationships in the context of the elasto-brittle damage model for rock matrix\textcolor{black}{,} (c) normal aperture versus normal stress\textcolor{black}{,} (d) shear stress (or shear dilation) versus shear displacement for pre-existing fractures.}}\label{fig:constitutive}
\end{figure}

We model the nonlinear normal opening and closure behaivour of rock fractures using an exponential function \citep{Rutqvist2002} (see Fig. \ref{fig:constitutive}c) :
\begin{equation}
b_\mathrm{n}=b_\mathrm{r}+(b_\mathrm{0}-b_\mathrm{r})\mathrm{exp}(\xi\sigma_\mathrm{n}),
\end{equation}
where $b_\mathrm{0}$ and $b_\mathrm{r}$ denote the initial and residual apertures, respectively; $\sigma_\mathrm{n}$ is the normal stress; $\xi = 1/[K_\mathrm{n0}(b_\mathrm{0}-b_\mathrm{r})]$ is the stress-aperture correlation
coefficient, with $K_\mathrm{n0}$ being the initial normal stiffness. The normal stiffness of the fracture changes in response to the normal stress, exhibiting a nonlinear behaviour as described by:
\begin{equation}
K_\mathrm{n}=\frac{\partial \sigma_\mathrm{n}}{\partial b_\mathrm{n}}=\frac{b_\mathrm{0}-b_\mathrm{r}}{b_\mathrm{n}-b_\mathrm{r}}K_\mathrm{n0}.
\end{equation}
The shear deformation of a fracture obeys Coulomb's friction law, expressed as \citep{Jaeger2007} (see Fig. \ref{fig:constitutive}d):
\begin{equation}
\tau_\mathrm{s}=\left\{
\begin{split}
K_\mathrm{s} u_\mathrm{s}, &\ u_\mathrm{s} < u_\mathrm{p} \\
\tau_\mathrm{p} \quad, &\ u_\mathrm{s} \geqslant u_\mathrm{p}\textcolor{black}{\  ,}
\end{split}
\right.
\end{equation}
In this context, $\tau_\mathrm{s}$ and $u_\mathrm{s}$ denote the shear stress and shear displacement, respectively; $K_\mathrm{s}$ is the fracture shear stiffness of the fracture; $\tau_\mathrm{p} = -\sigma^{'}_\mathrm{n} \mathrm{tan} \phi_\mathrm{f}$ denotes the peak shear stress, where $\phi_\mathrm{f}$ is the friction angle. The peak shear displacement is given by $u_\mathrm{p} = \tau_\mathrm{p}/K_\mathrm{s}$, which marks the displacement threshold beyond which the fracture starts to frictionally slide. Shear dilation is calculated in an incremental way as follows \citep{Lei2022} (see Fig. \ref{fig:constitutive}d):
\begin{equation}
\textcolor{black}{\mathrm{d}}v_\mathrm{s}=\left\{
\begin{split}
\mathrm{tan} \phi_\mathrm{d} \textcolor{black}{\mathrm{d}}u_\mathrm{s}, &\ 0 \leqslant u_\mathrm{s} \leqslant u_\mathrm{r} \\
0 \qquad, &\ u_\mathrm{s} \textgreater u_\mathrm{r} \textcolor{black}{\  ,}
\end{split}
\right.
\end{equation}
with $\phi_\mathrm{d}$ being the dilation angle and $u_\mathrm{r}$ the residual shear displacement beyond which dilation plateaus. The fracture aperture $b_\mathrm{f}$ under a combined normal and shear loadings is finally derived as \citep{Lei2016b,Lei2022}:
\begin{equation}
b_\mathrm{f}=b_\mathrm{n}+v_\mathrm{s}\label{b}.
\end{equation}

\subsection{Hydrogen diffusion and steel degradation}
Below we present the governing equations for modelling hydrogen diffusion as well as the constitutive equations for modelling HE. Several simplifications are made to mimic the hydrogen diffusion process in our finite element model. Firstly, the model considers only two types of sites: lattice sites (regular atomic positions) and trapping sites (locations where hydrogen can be trapped). This simplification reduces complexity by limiting the number of pathways that hydrogen atoms can move through. Furthermore, all saddle points, which represent intermediate states in the energy landscape for hydrogen movement, are assumed to have the same activation energy across the entire crystal. We simulate the hydrogen diffusion process and employ the hydrogen enhanced localised plasticity (HELP) mechanism to describe the plastic behaviour of the steel lining as hydrogen concentration increases \citep{Diaz2016}:
\begin{equation}
(\frac{\partial C_{T}}{\partial C_{L}} + 1)\frac{\partial C_{L}}{\partial t} - \nabla(D_{L} \nabla C_L + \nabla (\frac{D_{L} C_{L} \overline{V_{H}}}{R T} \nabla \sigma_{h}))
\end{equation}
where $C_{L}$ and $C_{T}$ correspond to the lattice and trapping concentration of hydrogen, respectively; $D_L$ is the lattice diffusion coefficient, and $\overline{V_{H}}$ denotes the partial molar volume; $R$ and $T$ stand for gas constant and temperature, which equal to 8.314 J/ (mol·K) and 293.15 K, respectively; $\sigma_{h}$ represents the hydrostatic stress.

Furthermore, Oriani's formulation that governs the equilibrium between lattice and trapping concentration \citep{ORIANI1970147} is adopted to solve ${\partial C_{T}}/{\partial C_{L}}$:
\begin{equation}
\frac{\theta_{T}}{1-\theta_{T}} = \frac{\theta_{L}}{1-\theta_{L}}K_{LT}
\end{equation}
where $\theta_{T}$ and $\theta_{L}$ stand for trapping and lattice occupancy, receptively, and $K_{LT}$ can be calculated from an Arrhenius law \citep{Diaz2016}:
\begin{equation}
K_{LT} = \mathrm{exp}(-\frac{E_{b}}{k_{B} T})
\end{equation}
where $E_{b}$ is the binding energy and $k_{B}$ is the Boltzmann constant ($1.380649 \times 10^{-23}$ J/K). By assuming $\theta_{L} \ll 1$, the evolution of hydrogen concentration can be solved as \citep{SOFRONIS1989317}:
\begin{equation}
\frac{\partial C_T}{\partial C_L} = \frac{C_T\theta_T}{C_L}
\end{equation}

If the effect of HE is absent, steel is governed by an elasto-plastic constitutive law with exponential hardening, with the yield stress given by \citep{Molavitabrizi2022}:
\begin{equation}
\sigma_\mathrm{ys} = \sigma_\mathrm{ys0}(1+\frac{\varepsilon^{p}}{\varepsilon_\mathrm{0}})^{N}
\end{equation}
where $\sigma_\mathrm{ys0}$ is the initial yield stress, $N$ is the hardening exponent, $\varepsilon^{p}$ is the equivalent plastic strain, and $\varepsilon_{0}$ is reference strain, which equals to $\sigma_\mathrm{ys0}/E_\mathrm{s}$ with $E_\mathrm{s}$ being the elastic modulus of steel.

In the presence of hydrogen, the yield stress of steel would reduce due to the effect of HE, which is modelled as \citep{Molavitabrizi2022}:
\begin{equation}
\sigma_\mathrm{ys}(C_{tot})=\left\{
\begin{split}
[(\xi_{s}-1)\frac{C_{tot}}{C_0}+1]\sigma_\mathrm{ys},  \quad &\sigma_\mathrm{ys}(C_{tot}) \textgreater \Psi \sigma_\mathrm{ys}  \\
\Psi \sigma_\mathrm{ys} \qquad  \qquad,  \quad & \sigma_\mathrm{ys}(C_{tot}) \leqslant \Psi \sigma_\mathrm{ys} \  ,
\end{split}
\right.
\end{equation}
where $\xi_{s}$ is a softening parameter, $\Psi \sigma_\mathrm{ys}$ denotes the the lowest value of the yield stress, and ${C_0} = 1$ is the reference concentration. Once yielded, the steel is considered to be degraded.

\section{Model setup}
\label{sec:Model setup}

\begin{figure}[!htb]
\centering
\includegraphics[width = 14 cm]{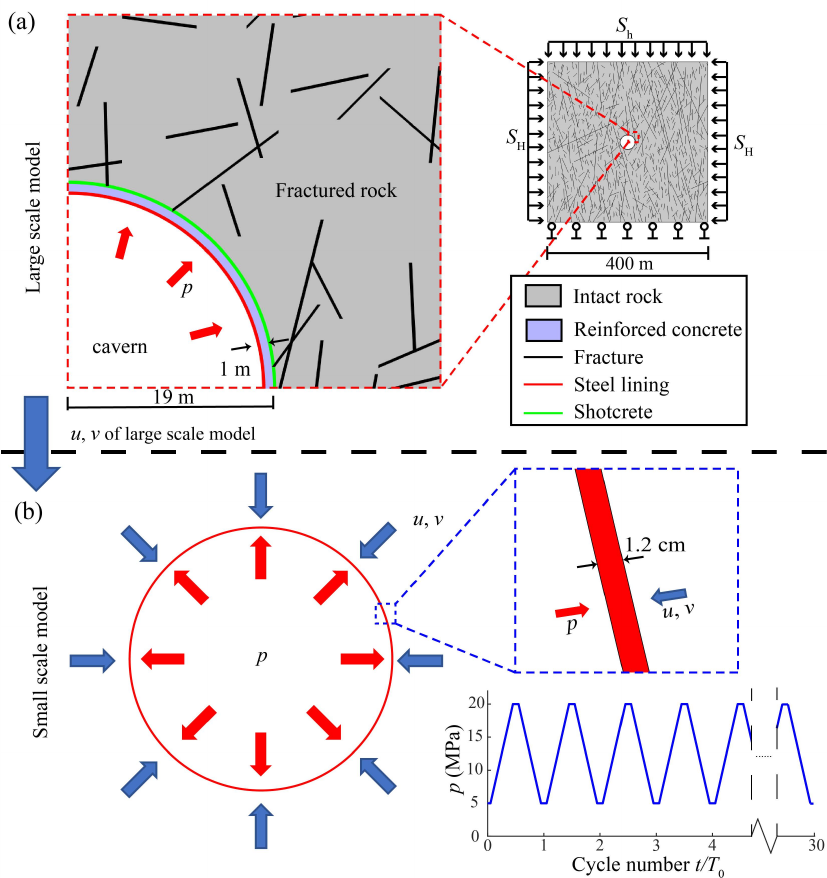}
\par
\caption{Model design and boundary condition of the multiscale model, including (a) a large-scale model representing an LRC situated in a fractured rock mass and (b) a small-scale model capturing the response of the steel lining subject to cyclic internal pressurisation and boundary displacement constraints, as well as hydrogen diffusion.}\label{fig:model setup}
\end{figure}

We adopt a multiscale modelling strategy and build the model in COMSOL Multiphysics \citep{COMSOL2018} (Fig. \ref{fig:model setup}). In this framework, a large-scale model represents an LRC situated within a fractured rock mass, while a small-scale model specifically calculates the hydrogen diffusion and embrittlement within the steel lining of the LRC. The two models are coupled at the interface between the steel lining and the concrete layer with displacement compatibility imposed.

We construct the model based on the available data of the LRC demonstration plant at Skallen, Sweden \citep{Glamheden2006}. Material properties for the fractured rock mass are given in Table \ref{table:material1}. The LRC of 38 m in diameter consists of a steel lining of 1.2 cm thickness, a reinforced concrete layer of 1 m thickness, and a shortcrete layer in contact with the surrounding rock mass \citep{Glamheden2006}. The material properties of the steel and concrete are given in Table \ref{table:material2}. The reinforced concrete lining assumes the C30/37 concrete properties in our simulation. Reinforcement consists of steel bars measuring 1.6 centimeters in diameter spaced at intervals of 15 centimeters. The influence of reinforcement is accounted for by adopting an increased modulus and incorporating a residual strength for the concrete layer. The steel lining and shortcrete layer are modelled using interface elements rather than volumetric elements due to their thin nature. The steel lining is specified as AISI 4135. High-strength steels are used here for two reasons: (i) they increase the gas filling threshold, allowing the LRC to reach more critical scenarios, and (ii) they are more susceptible to hydrogen embrittlement as susceptibility to this phenomenon increases with strength \citep{Meda2023}. Our selected value of $10^{-9} \ \mathrm{m^2/s}$ for lattice diffusion coefficient falls in the typical range of around $10^{-9}$ to $10^{-8}\ \mathrm{m^2/s}$ reported in the literature \citep{SOFRONIS1989317,KROM1999971}. Partial molar volume is selected based on \cite{SOFRONIS1989317} and \cite{KROM1999971}. The softening parameter $\xi_s$ and lowest limit $\Psi$ listed in Table \ref{table:material2} are obtained from \cite{Ahn2007}. The shortcrete layer has a cohesion of 2 MPa and a friction coefficient of 1 \citep{Damasceno2023a}.

\begin{figure}[!htb]
\centering
\includegraphics[width= 14 cm]{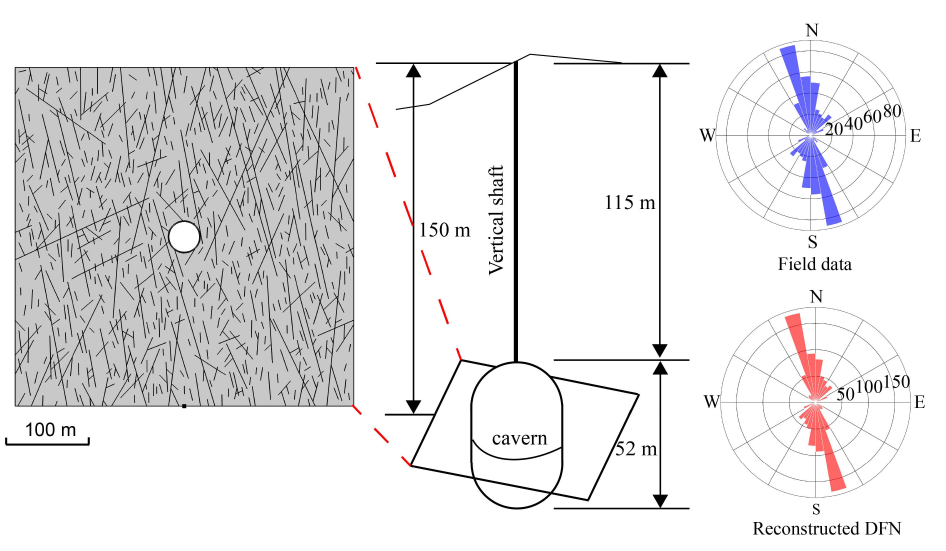}
\par
\caption{The discrete fracture network generated based on the site characterisation data of the LRC demonstration plant at Skallen, Sweden \cite{Glamheden2006}, together with a comparison of fracture orientation between field data and numerical model.}\label{fig:FracInfo}
\end{figure}

This LRC is located 150 meters below ground surface (Fig. \ref{fig:FracInfo}) with the in-situ stress state defined according to the field condition at the Skallen site \citep{Glamheden2006}, where the minimum horizontal stress is $S_\mathrm{h} = -8.8 \ \mathrm{MPa}$ and the maximum horizontal stress is $S_\mathrm{H} = -4.4 \ \mathrm{MPa}$. The in-situ stresses are applied orthogonally to the boundaries of the large-scale model, which represents a 2D horizontal cross-section of the system. We choose an extended domain size of 400 m to minimise potential boundary effects. The discrete fracture network approach is employed to represent the distribution of natural fractures in rock, according to the site characterisation data at Skallen \citep{Glamheden2006}. Here, the fracture lengths follow a power-law distribution, while the fracture orientations are generated based on geological mapping data from the surrounding area, as shown in the right stereonets of Fig. \ref{fig:FracInfo}. In the small-scale model, the outer boundary of the steel lining is informed by the displacement derived from the large-scale model, while the inner boundary of the steel lining is subject to gas pressure and hydrogen concentration. Cyclic gas loading is applied, with each cycle lasting for one day ($T_\mathrm{0}$ = 24 h). The initial cavern pressure is set at 5 MPa and incrementally increased over time in a linear manner until it reaches 20 MPa. During each cycle, the pressure stabilises at both initial and peak values for 0.1 days. The concentration of hydrogen $C_0$ at the inner boundary of the steel lining is calculated using Sievert's law \citep{Koren2023}:
\begin{equation}
C_0 = \Omega \sqrt{p\cdot \mathrm{exp}(\frac{pB}{RT})} 
\end{equation}
where $B = 1.584 \times 10^{-5} \mathrm{\ m^3\cdot mol^{-1}}$, and $\Omega$ is Sievert's constant that equals to 0.0125 $\mathrm{wppm \cdot bar^{-1/2}}$ \citep{Koren2023}.

In the large-scale model, the concrete layer was discretised into structured quadrilateral elements with an average size of 0.2 m. The rock matrix was meshed into unstructured triangular elements that vary in size, starting from 0.2 m near the LRC boundary and growing up to 5.0 m at the far end of the domain. Steel lining and shortcrete were modelled using linear elements, while fractures were represented by joint elements placed between adjacent triangular elements. This leads to approximately 300,000 finite elements in our large-scale model. In the small-scale model, the steel lining was discretised using structured quadrilateral elements with a radial division into 5 layers and a tangential division into 3,000 segments.

\begin{table}[!htb]
\centering
\caption{\textcolor{black}{Material properties of fractured rock masses.}}
\label{table:material1}
\begin{tabular}{ l l l l}  
\hline
&Material properties& Values & Units \\ 
\hline

 \multirow{6}{*}{Rock}&Density $\rho_\mathrm{r}$ & 2650 & $\mathrm{kg/m^3}$ \\ 
&Young's modulus $E_\mathrm{r}$  & 70 & $\mathrm{GPa}$ \\ 
&Poisson's ratio $\nu_\mathrm{r}$  & 0.25 & $\mathrm{-}$ \\
&Tensile strength $f_\mathrm{rt}$  & 10 & $\mathrm{MPa}$ \\
&Compressive strength $f_\mathrm{rc}$  & 300 & $\mathrm{MPa}$ \\
&Residual strength ratio $\eta$ & 0.1 & $\mathrm{-}$ \\
\hline

 \multirow{6}{*}{{Fractures}}&Initial normal stiffness $K_\mathrm{n0}$ & 50 & $\mathrm{GPa/m}$ \\ 
&Shear stiffness $K_\mathrm{s}$ & 10 & $\mathrm{GPa/m}$ \\ 
&Friction angle $\phi_\mathrm{f}$ & 30 & \textcolor{black}{$\mathrm{^\circ}$} \\ 
&Dilation angle $\phi_\mathrm{d}$ & 3 & \textcolor{black}{$\mathrm{^\circ}$} \\ 
&Residual displacement $u_\mathrm{r}$ & 3 & $\mathrm{mm}$ \\
&Initial aperture $b_{0}$ & 0.1 & $\mathrm{mm}$ \\
&Residual aperture $b_{\mathrm{r}}$ & 0.01 & $\mathrm{mm}$ \\
\hline
\end{tabular}
\end{table}

\begin{table}[!htb]
\centering
\caption{Material properties of concrete and steel.}
\label{table:material2}
\begin{tabular}{ l l l l}  
\hline
Material properties& Values & Units \\ 
\hline
 \multirow{10}{*}{\makecell[l]{Reinforced\\concrete\\lining}} 
&Density $\rho_\mathrm{c}$ & 2500 & $\mathrm{kg/m^3}$ \\ 
&Young's modulus $E_\mathrm{c}$  & 34 & $\mathrm{GPa}$ \\ 
&Poisson's ratio $\nu_\mathrm{c}$  & 0.2 & $\mathrm{-}$ \\
&Elastic limit of tensile strain $\varepsilon_\mathrm{ct0}$  & $1 \times 10^{-4}$ & $\mathrm{-}$ \\
&Elastic limit of compressive strain $\varepsilon_\mathrm{cc0}$  & $1 \times 10^{-3}$ & $\mathrm{-}$ \\
&Mazars damage coefficient $A_\mathrm{t}$ & 0.7 & $\mathrm{-}$ \\
&Mazars damage coefficient $A_\mathrm{c}$ & 1.12 & $\mathrm{-}$ \\
&Mazars damage coefficient $B_\mathrm{t}$ & 7000 & $\mathrm{-}$ \\
&Mazars damage coefficient $B_\mathrm{c}$ & 1000 & $\mathrm{-}$ \\
&Maximum damage $\omega_\mathrm{cmax}$ & 0.95& $\mathrm{-}$ \\
\hline
 \multirow{12}{*}{\makecell[l]{Steel\\lining}} 
&Density $\rho_\mathrm{s}$ & 7850 & $\mathrm{kg/m^3}$ \\ 
&Young's modulus $E_\mathrm{s}$ & 195 & $\mathrm{GPa}$ \\ 
&Poisson's ratio $\nu_\mathrm{s}$ & 0.3 & $\mathrm{-}$ \\ 
&Yield strength $\sigma_\mathrm{sy0}$ & 1320 & $\mathrm{MPa}$ \\ 
&Hardening exponent $N$ & 0.11 & $-$ \\ 
&Softening parameter $\xi_s$ & 0.99 & $-$ \\ 
&Lowest limit $\Psi$ & 0.5 & $-$ \\ 
&Lattice diffusion coefficient $D_L$ & 1e-9 & $\mathrm{m}^2/\mathrm{s}$ \\
&Partial molar volume $\overline{V_H}$ & 2e-6 & $\mathrm{m^3/mol}$ \\
&Lattice site concentration $N_L$ & 1e6 & $\mathrm{mol/m^3}$ \\
&Trap site concentration $N_T$ & 20 & $\mathrm{mol/m^3}$ \\
&Trap binding energy $E_b$ & 30e3 & $\mathrm{J/mol}$ \\

\hline
\end{tabular}
\end{table}

\begin{figure}[!htb]
\centering
\includegraphics[width=\textwidth]{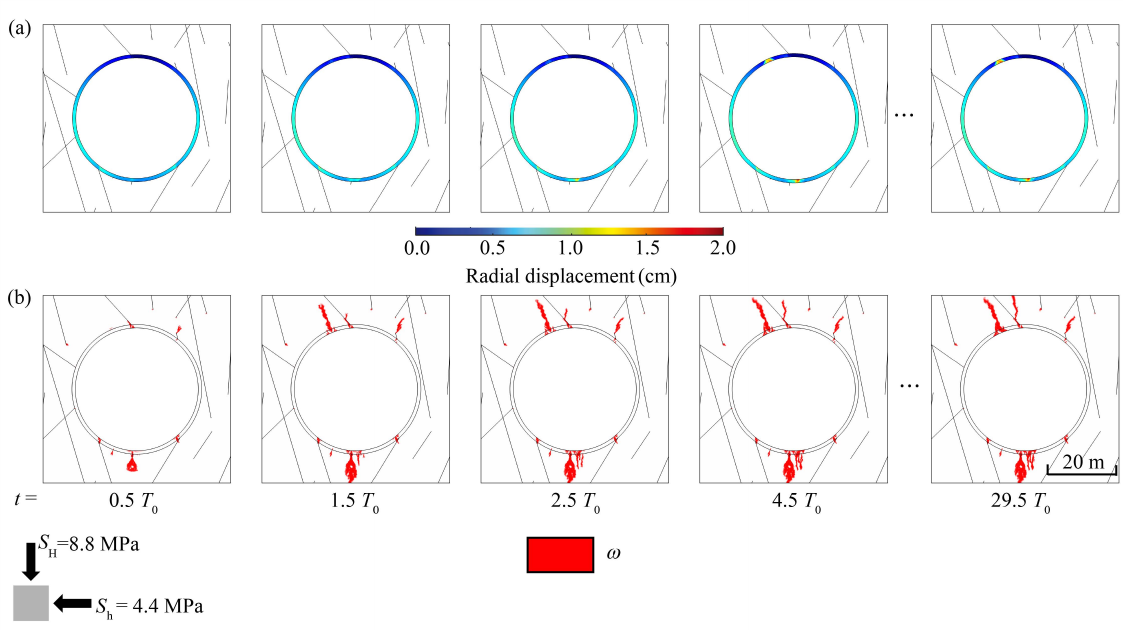}
\par
\caption{Distribution and evolution of (a) radial displacement in the concrete lining and (b) damage in concrete and rock over multiple loading cycles.}\label{fig:DmgCloud}
\end{figure}

\begin{figure}[!htb]
\centering
\includegraphics[width= 14 cm]{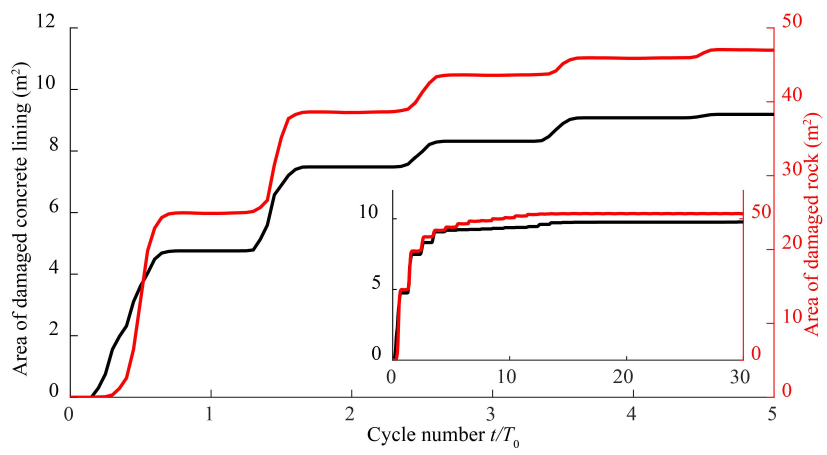}
\par
\caption{Temporal evolution of damaged area in concrete and rock mass during hydrogen gas filling-emptying cycles.}\label{fig:LargeScaleDmgLine}
\end{figure}

\begin{figure}[!htb]
\centering
\includegraphics[width= 14 cm]{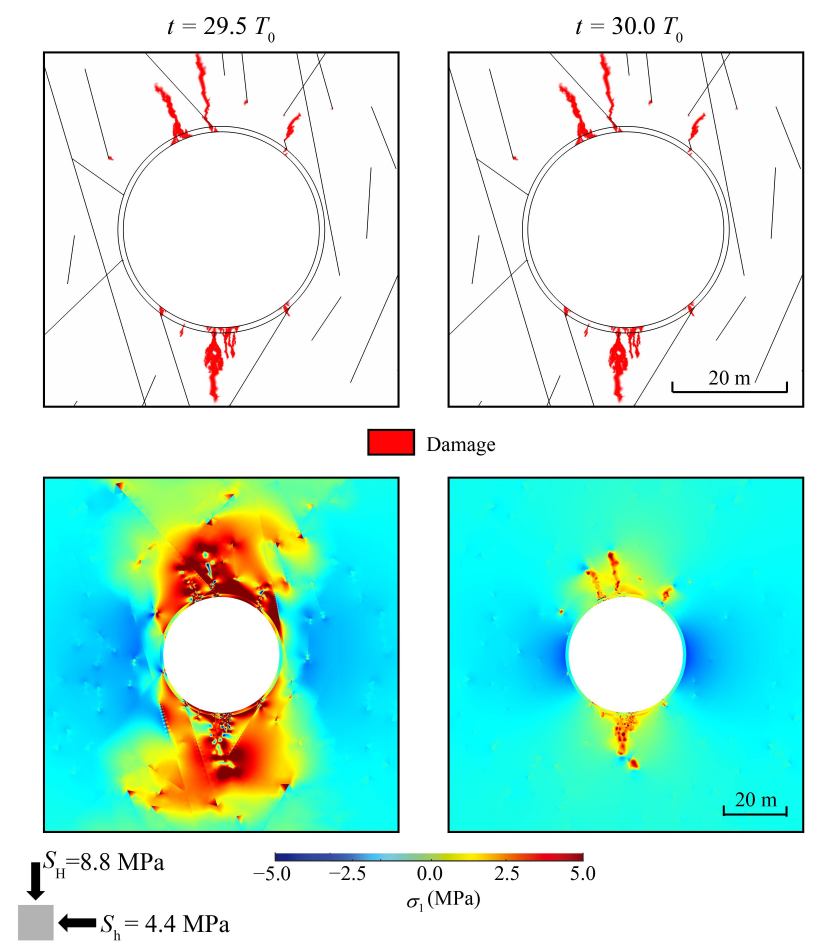}
\par
\caption{Distribution of damage and local maximum principal stress in the vicinity of the LRC under a peak (left panel) and residual (right panel) internal pressure.}\label{fig:DmgStressCloud}
\end{figure}

\begin{figure}[!htb]
\centering
\includegraphics[width= 14 cm]{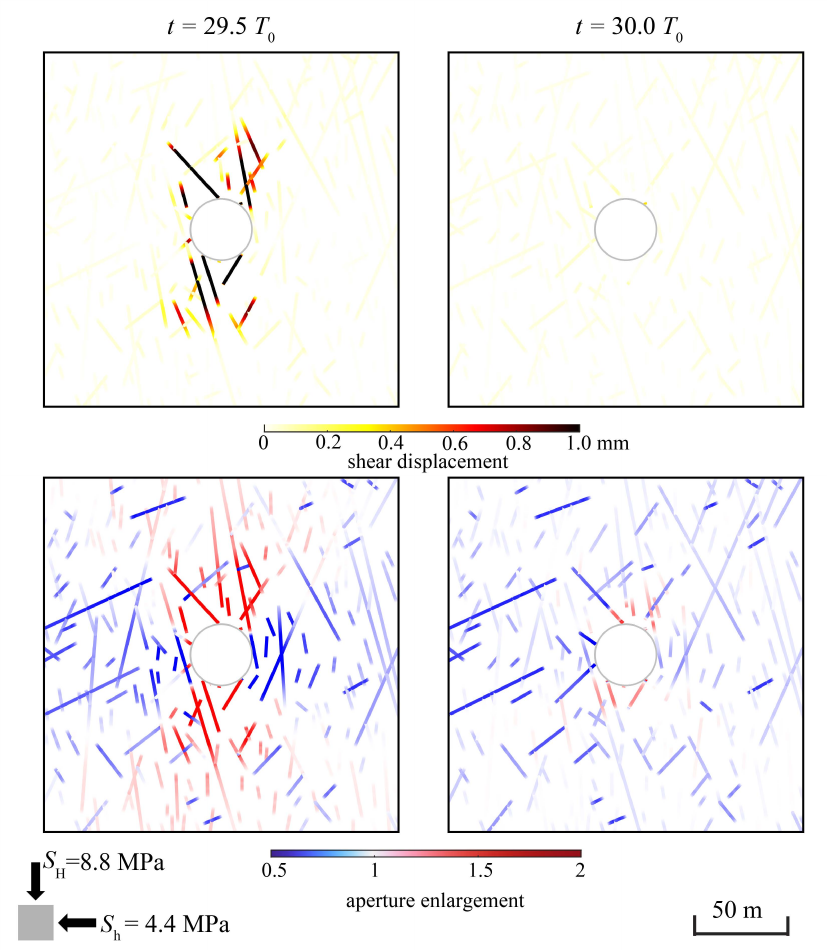}
\par
\caption{Shear displacement and aperture variation in cavern vicinity under initial and peak pressure.}\label{fig:usap}
\end{figure}

\section{Simulation results}
\label{sec:simulation results}

\subsection{Deformation and damage evolution in the rock mass and concrete lining}
\label{sec:results of large scale model}

Fig. \ref{fig:DmgCloud} shows the spatial distribution and temporal evolution of radial displacement in the LRC lining as well as the damage patterns in rock and concrete. Note that we adopt the solid mechanics sign convention with outward expansion being positive and inward contraction being negative. Variations in the displacement pattern of the LRC lining are pronounced around pre-existing fractures, highlighting their strong influence on LRC deformations (Fig. \ref{fig:DmgCloud}a). Around the cavern, predominant tensile cracks developed along the maximum horizontal stress direction. Damage in rock is also characterised by wing cracks originating from the tips of pre-existing fractures. In the concrete layer, damage is observed in the sections where the concrete is subject to a tensile loading and intersects with either pre-existing fractures or newly formed cracks in the surrounding rock mass. During the initial cycles ($t \leqslant 5 T_0$), both the concrete lining and rock mass experience significant increases in damage under cyclic loading. However, after this initial stage, further changes in damage are negligible. A strong correlation can be observed between the displacement field of the LRC lining and the damage distribution in the system. Notably, the damage occurring at the front and back sections of the lining is accompanied with significant LRC deformation in these areas. The temporal evolution of damaged areas within both the rock mass and concrete lining are further illustrated in Fig. \ref{fig:LargeScaleDmgLine}. There is a noticeable repeating pattern over cycles, with damage generally prevailing at the beginning of each cycle. Most damage occurs in the early cycles and levels off after around ten cycles.

In Fig. \ref{fig:DmgStressCloud}, we analyse the distribution of damage and local maximum principal stresses in the vicinity of the cavern under the peak and residual gas pressures. The regions exhibiting high maximum principal stresses are areas with concentrated tensile stresses, which align with the anticipation from the classical Kirsch's solution \citep{Jaeger2007}. Significant stress heterogeneity is observed in the fractured rock when the cavern is loaded by the peak gas pressure (e.g. $t = 29.5\ T_0$), resulting from the complex interplay of cavern loading, rock mass deformation, and in-situ stresses. When the internal pressure is reduced to the residual value (e.g. $t = 30\ T_0$), the stress heterogeneity becomes much less, although some high tensile stress still remain locally, especially in the regions where new cracks are formed. By examining the damage pattern and stress distribution, we can see that tensile stress concentration plays a crucial role in generating damage in both concrete and rock, while compression-induced damage seems to be minor.

Fig. \ref{fig:usap} illustrates how gas pressurisation within the LRC can influence the shear and normal deformation of pre-existing fractures around the LRC. In this figure, the upper panel shows the pattern of fracture shear displacement after cyclic loading. The lower panel gives the variation in fracture aperture quantified by the ratio of fracture apertures before and after the cavern pressurisation. A ratio greater than 1 (highlighted in red) indicates that the aperture is enlarged, while a ratio less than 1 (highlighted in blue) signals decreased aperture. Gas infilling induces significant shear deformation along pre-existing fractures in the front and back regions of the LRC, which correlates well with the tensile stress field simulated (e.g. Fig. \ref{fig:DmgStressCloud}). Fractures in these regions also experience significant opening driven by shear dilation, while those in other regions show minimal reactivation during gas pressurisation. The rock mass undergoes an overall compaction as a result of the expansion of the LRC structure, leading to widespread closure of fractures in the rock mass.

\begin{figure}[!htb]
\centering
\includegraphics[width= 14 cm]{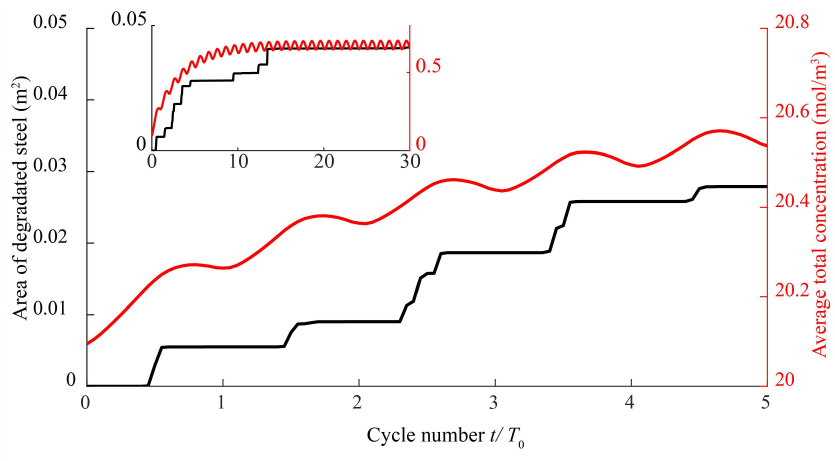}
\par
\caption{Temporal evolution of degradated area (in black) and hydrogen concentration (in red) in the steel lining where degradation accumulates in stages and hydrogen concentration experiences increased volatility.}\label{fig:CLandSteelDmg}
\end{figure}

\begin{figure}[!htb]
\centering
\includegraphics[width= 14 cm]{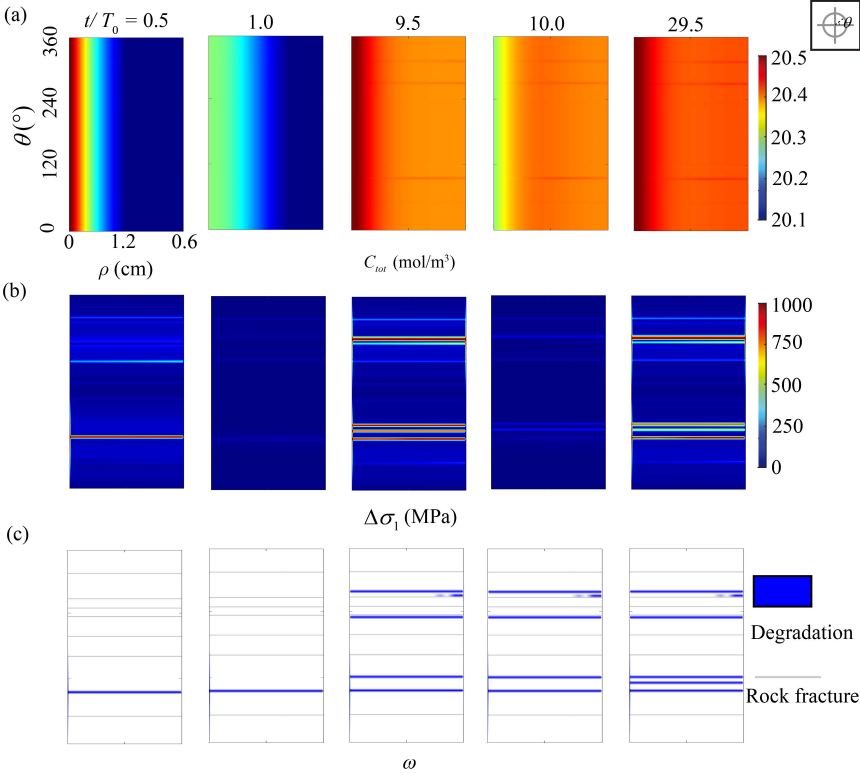}
\par
\caption{Spatial distribution of (a) hydrogen concentration, (b) maximum principal stress variation, and (c) strength degradation in the steel lining at different loading stages  ($t/T_0$), where variations are observed at similar locations across all three rows of plots.}\label{fig:SteelCloud}
\end{figure}

\begin{figure}[!htb]
\centering
\includegraphics[width= 14 cm]{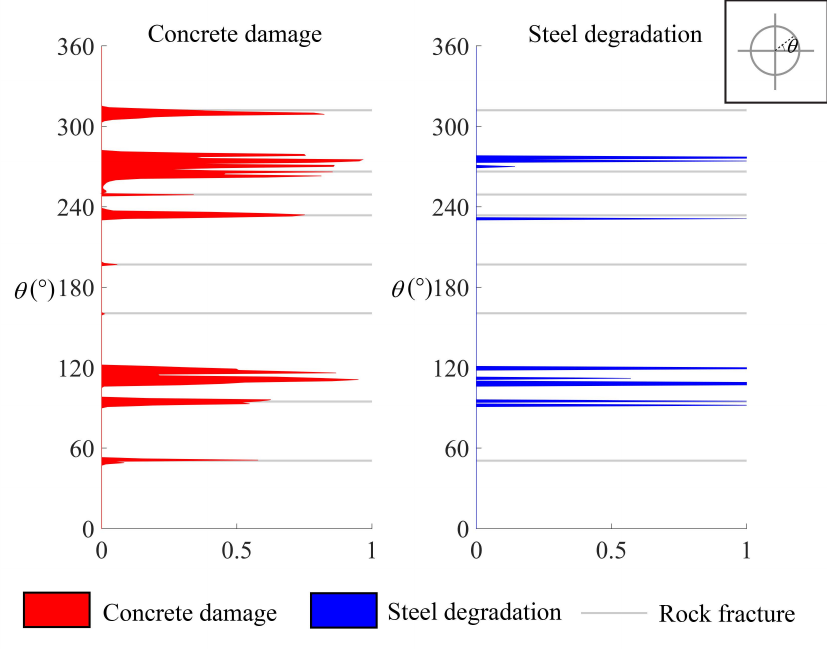}
\par
\caption{Cumulative damage in the concrete layer and cumulative degradation in the steel lining, with a comparison to the position of rock fractures intersecting with the LRC.}\label{fig:SteelandConcreteDmgLine}
\end{figure}

\begin{figure}[!htb]
\centering
\includegraphics[width= 14 cm]{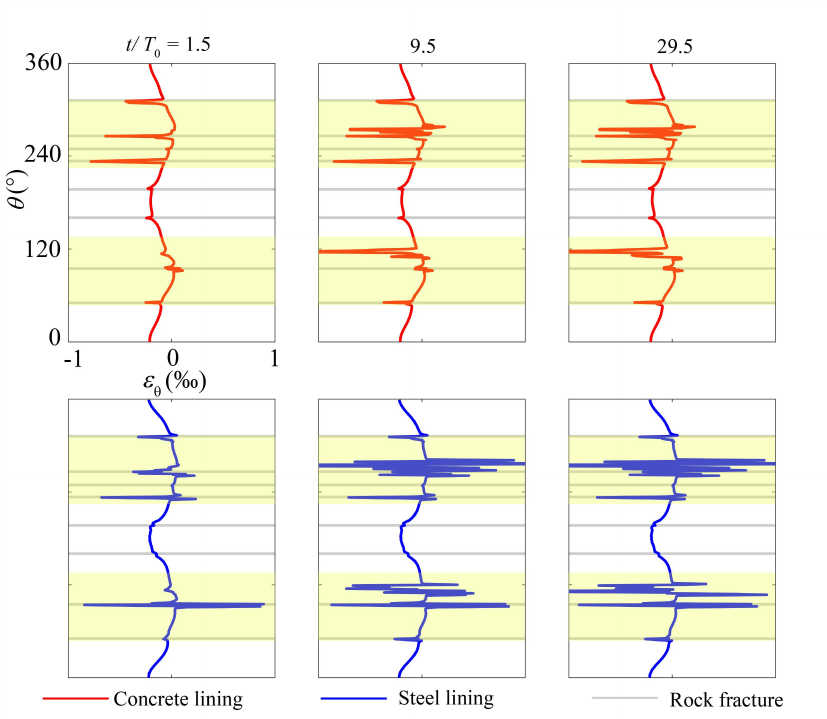}
\par
\caption{Distribution and evolution of tangential strain in the concrete and steel linings at different loading stages.}\label{fig:SteelandConcreteTanStrain}
\end{figure}

\begin{figure}[!htb]
\centering
\includegraphics[width= \textwidth]{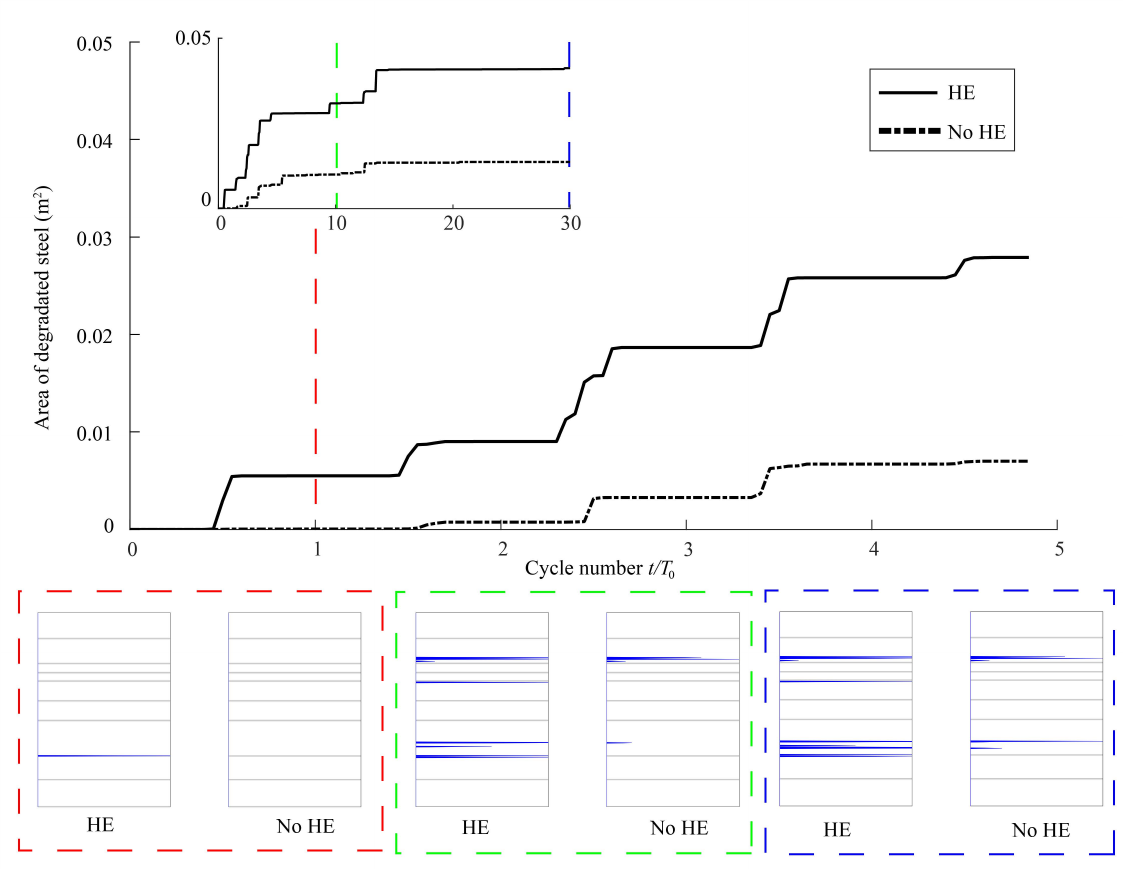}
\par
\caption{Temporal evolution of degradated area in the steel lining with and without hydrogen embrittlement (HE), together with the degradation pattern in the steel lining at $t/T_0 = 1,10$ and $30$ (from left to right in the lower panel).}\label{fig:HEandSol}
\end{figure}

\subsection{Hydrogen accumulation and strength degradation in the steel lining}
\label{sec:results of small scale model}

Fig. \ref{fig:CLandSteelDmg} shows the temporal evolution of degraded area and hydrogen concentration in the steel lining. The hydrogen concentration is represented by the red line, which exhibits a cyclically growing pattern in the initial cycles, followed by a stabilisation with minor fluctuations in the later stages. The steel degradation area continues to grow even after 15 loading cycles, while both the hydrogen concentration (shown by the red line in Fig. \ref{fig:CLandSteelDmg}) and the damage to the concrete lining and rock (depicted in Fig. \ref{fig:LargeScaleDmgLine}) remain generally unchanged. This result indicates that HE has a prolonged impact on the LRC's performance, leading to fatigue-type degradation of the steel lining over time.

In Fig. \ref{fig:SteelCloud}, we further show the spatial distribution and temporal evolution of hydrogen concentration, maximum principal stress, steel degradation in the steel lining. The horizontal axis represents the thickness $\rho$ of the steel lining, while the vertical axis denotes the central angle, ranging from 0 to 360$^\circ$ along the tangential direction of the lining. Most of the variation in hydrogen concentration occurs in the thickness direction, with minimal change in concentration observed as the angle varies at the same depth $\rho$. This suggests that the overall concentration is primarily influenced by the boundary concentration level. In addition to the gradient along the thickness direction, local fluctuations are noticeable for $t/T_0 = 9.5, 10, 29.5$. These fluctuations align closely with changes in stress and degradation, indicating that steel degradation influences the local hydrogen concentration. In the second row of this figure, we plot the maximum principal stress variation, which show a good consistency with the degradation pattern. For the degradation evolution shown in the third row of this figure, the horizontal grey lines mark the positions where pre-existing fractures intersect with the LRC lining. Two types of failure patterns can be observed in the degradation field. The degradation near $\theta = 300^\circ$ and $240^\circ$ is possibly related to nearby fractures in rock, where stress concentration occurs due to shear dislocation along those pre-existing fractures. For the failure around $\theta = 120^\circ$ associated with no nearby pre-existing fractures in rock, it seems to be governed by the deformation field of the LRC under the in-situ stress and cavern pressurisation (see Fig. \ref{fig:DmgCloud}). This second type of degradation develops in a progressive manner, increasing as the number of cycles grows, which highlights the interaction between HE in the steel lining and variation of local stress field under cyclic loading.

Fig. \ref{fig:SteelandConcreteDmgLine} illustrates the variation in concrete damage and steel degradation  around the cavern. A strong correlation is observed between concrete damage and fracture intersections, where a fracture that intersects the lining promotes concrete damage. Notably, fractures that intersect with the lining in the tensile loading zone (45° < $\theta$ < 135° and 225° < $\theta$ < 315°) lead to more significant damage evolution nearby, while fractures in the compressive regions cause only minor damage. A similar pattern is observed in the steel degradation field, where regions near fractures are more susceptible to strength degradation, and significant yielding occurs primarily in the tensile zone. This indicates that pre-existing fractures influence not only the concrete lining but also the inner steel tank. One key difference between concrete damage and steel degradation with HE is that steel degradation is more localised, whereas concrete damage is more diffuse. Additionally, for the concrete lining, at all the fracture-cavern intersections, damage can be observed, although of variable extents. In contrast, at only certain interactions, steel degradation is seen, reflecting the protection effect of the outer concrete lining on the inner steel lining.

In Fig. \ref{fig:SteelandConcreteTanStrain}, variations in tangential strain for both the concrete (represented by the red line) and steel linings (represented by the blue line) are shown. The yellow-shaded areas indicate the LRC sections under tensile loading. It is clear that the changes in tangential strain for both materials are influenced by the in-situ stress conditions and the spatial distribution of pre-existing fractures. The overall tangential strain pattern is primarily controlled by the in-situ stress, with tensile strains occurring at the front and back sections of the lining (i.e. $45^\circ<\theta<135^\circ$ and $225^\circ<\theta<315^\circ$), while the remaining sections are dominated by compressive strains. Notably, significant concentrations of tensile strain are observed in both the steel and concrete linings near some pre-existing fractures (reflected by the spikes in Fig. \ref{fig:SteelandConcreteTanStrain}). At those fracture-cavern intersections, strain fluctuations are noticeable, but sharp increases in tangential strain are only observed in fractures located within the tensile regions. Although both concrete and steel linings exhibit a nonuniform strain distribution, the magnitude of tensile strain concentration is considerably higher in the steel lining than that in concrete. Additionally, for the steel lining, while strain variations are seen at all fracture-cavern intersections, strong fluctuations occur in the later stages at some specific locations such as at $\theta \approx 240^\circ$ and $\theta \approx 360^\circ$ when $t/T_0 > 1.5$, where strong degradation develops.

\section{Discussion}
\label{sec:Discussion}

To gain deeper insights into the LRC failure behaviour and mechanism as well as its interaction with the hydrogen diffusion process, we conduct further investigations to model the response of an LRC system with no HE effect and compare with our results considering the HE effect. In Fig. \ref{fig:HEandSol}, we show the temporal evolution of degradated area in model with and without HE. For the case without HE, fewer instances of degradation are observed during cavern loading, which generally occur in the first 10 cycles, while the model with HE experience much stronger and more sustained degradation, suggesting that the HE effect can indeed significantly affect the LRC performance. In the lower part of Fig. \ref{fig:HEandSol}, we also compare the two models regarding the accumulated degradation along the steel lining at different loading stages, highlighting two key mechanisms that explain the differences between the two models. Firstly, near the fracture at $\theta = 240^\circ$, the model without HE shows no failure, whereas failure occurs in the model with HE, indicating that the presence of hydrogen diffusion amplifies the impact of pre-existing fractures on the LRC structure. Additionally, at the section around $\theta = 120^\circ$, both models exhibit degradation, but the progression is more pronounced and faster in the model with HE. It can be attributed to the coupling between the concentrations of hydrogen and stress in the steel lining. More specifically, HE leads to the redistribution of stress within the steel, where stress concentration occurs in certain areas where hydrogen atoms would further accumulate. This accumulation increases the local hydrogen concentration, making the steel weaker by reducing its yield stress. As a result of this positive feedback, the steel becomes more susceptible to degradation during cyclic loading. Our simulation results highlight the critical impact of HE on the LRC integrity and performance.

Previous analytical studies \citep{johansson1995,Damasceno2023b} and numerical simulations \citep{Lu1998,Tunsakul2014,Perazzelli2016,Damasceno2023a} have highlighted that the stability of a pressurised cavern is primarily influenced by tensile cracking or damage in the surrounding rock mass. Other simulations have also shown that the presence and distribution of pre-existing fractures play a crucial role in the structural deformation of LRCs \citep{ZHAO2025252}. Our current simulation results further demonstrate that pre-existing fractures in rock can also affect the inner steel lining in the context of hydrogen storage, potentially jeopardising the gas-tight environment of the cavern due to hydrogen-induced fatigue failure. In cases where the LRC is situated within a rock mass containing numerous pre-existing fractures, the structural deformation of the cavern is characterised by damage in the concrete lining at various locations, which can lead to hydrogen accumulation and further steel degradation. Our findings suggest that fractures intersecting the LRC significantly influence the development of concrete damage, steel degradation, and the evolution of tangential strain in both the steel and concrete linings. Furthermore, our simulations indicate that the growth of new cracks in the concrete lining can result in localised degradation in the steel lining.

It is important to note that some aspects were not considered in the current study and will need to be addressed in future research. One aspect is related to the presence of groundwater, where overpressure and underpressure could be generated around the cavern during the LRC filling-emptying cycles. To investigate the evolution of pore water pressure and its effects on LRC performance, a fully-coupled hydro-mechanical model \citep{Zhao2021,Zhao2022} can be employed. Secondly, apart from the hydrogen-induced fatigue controlling the gas-tightness of the cavern, the reinforced concrete and fractured rock masses can also display time-dependent, mechanically-driven fatigue behaviour, posing threat to the structural integrity of the LRC system. Considering the creep behavior of the rock/concrete material is crucial when assessing the long-term performance of LRCs, which will be explored in our future work. Furthermore, more sophisticated models with hydrogen trapping and other degradation mechanisms \citep{Yu2024}, such as the Hydrogen-Enhanced Decohesion and Hydrogen
Enhanced Strain-Induced Vacancy mechanisms, may also need to be considered and explored.

\section{Conclusions}
\label{sec:conslusions}
In this paper, we developed a novel 2D multiscale model to simulate the behaviour of a lined rock cavern situated in fractured rock masses under cyclic hydrogen loading. Our simulations demonstrate that the damage in the concrete layer and rock masses as well as the degradation in the steel lining are strongly influenced by the presence of pre-existing fractures in rock and the in-situ stress conditions. Notably, more severe concrete damage and steel degradation are observed where fractures intersect with the extensionally loaded sections of the LRC. Our results reveal the existence of a strong positive feedback between the hydrogen concentration and stress redistribution in the steel lining. HE causes a redistribution of stress within the steel, elevating the stress in regions where hydrogen atoms accumulate. This in turn leads to an increased local concentration of hydrogen, which further weakens the steel by lowering its yield stress. Consequently, the steel becomes more vulnerable to degradation under cyclic loading. These findings highlight the significant impact of HE on the safety and performance of LRCs. Additionally, a comparison between models with and without HE reveals that hydrogen concentration plays a significant role in steel degradation, particularly in the context of the long-term LRC operation. Our research indicate that HE in LRCs can pose significant operational risks, particularly given the presence of fracture networks in rock masses. It is thus essential to consider these factors when assessing the safety and performance LRCs for sustainable underground hydrogen storage in fractured rocks.

\section*{Declaration of competing interest}

The authors declare that they have no known competing financial interests or personal relationships that could have appeared to influence the work reported in this paper.

\section*{Data availability }

\textcolor{black}{The data that support the findings of this study are available from the corresponding author upon reasonable request.}

\section*{Acknowledgements}
\label{sec:acknowledgements}
\textcolor{black}{Q.L. and H.Y.} are grateful for the support from the Nordic Energy Research (Grant No. 187658). \textcolor{black}{Z.Z.} thanks for the support from the National Natural Science Foundation of China (Grant No. 42377146).

\appendix
%\section{Appendix}

\printcredits

\bibliographystyle{model3-num-names}

\bibliography{cas-sc-template}


%Author biography without author photo.
%Author biography. Author biography. Author biography.
%Author biography. Author biography. Author biography.
%Author biography. Author biography. Author biography.
%Author biography. Author biography. Author biography.
%Author biography. Author biography. Author biography.
%Author biography. Author biography. Author biography.
%Author biography. Author biography. Author biography.
%Author biography. Author biography. Author biography.
%Author biography. Author biography. Author biography.

%\bio{figs/pic1}
%Author biography with author photo.
%Author biography. Author biography. Author biography.
%Author biography. Author biography. Author biography.
%Author biography. Author biography. Author biography.
%Author biography. Author biography. Author biography.
%Author biography. Author biography. Author biography.
%Author biography. Author biography. Author biography.
%Author biography. Author biography. Author biography.
%Author biography. Author biography. Author biography.
%Author biography. Author biography. Author biography.

%\bio{figs/pic1}
%Author biography with author photo.
%Author biography. Author biography. Author biography.
%Author biography. Author biography. Author biography.
%Author biography. Author biography. Author biography.
%Author biography. Author biography. Author biography.

\end{document}